\documentclass{sig-alternate-2013}

\newfont{\mycrnotice}{ptmr8t at 7pt}
\newfont{\myconfname}{ptmri8t at 7pt}
\let\crnotice\mycrnotice%
\let\confname\myconfname%

\permission{Permission to make digital or hard copies of all or part of this work for personal or classroom use is granted without fee provided that copies are not made or distributed for profit or commercial advantage and that copies bear this notice and the full citation on the first page. Copyrights for components of this work owned by others than the author(s) must be honored. Abstracting with credit is permitted. To copy otherwise, or republish, to post on servers or to redistribute to lists, requires prior specific permission and/or a fee. Request permissions from permissions@acm.org.}
\conferenceinfo{HT'14,}{September 1--4, 2014, Santiago, Chile. \\
{\mycrnotice{Copyright is held by the owner/author(s). Publication rights licensed to ACM.}}}
\copyrightetc{ACM \the\acmcopyr}
\crdata{978-1-4503-2954-5/14/09\ ...\$15.00.\\
http://dx.doi.org/10.1145/2631775.2631799
}

\makeatletter
\newif\if@restonecol
\makeatother

\usepackage[boxed]{algorithm2e}

\usepackage{url} 
\usepackage{graphicx}
\usepackage{tabularx}
\usepackage{subfigure}
\usepackage{balance}

\begin{document}

\title{The Shortest Path to Happiness: \\ Recommending Beautiful, Quiet, and Happy Routes in the City}
\numberofauthors{3}

\author{
\alignauthor
Daniele Quercia\\
       \affaddr{Yahoo Labs}\\
       \affaddr{Barcelona, Spain}\\
       \email{dquercia@yahoo-inc.com}
\alignauthor
Rossano Schifanella\titlenote{This work has been done while the author was visiting Yahoo Labs, Barcelona, within the framework of the Faculty Research and Engagement Program.}\\
	\affaddr{University of Torino}\\
	\affaddr{Torino, Italy}\\
       \email{schifane@di.unito.it}\\
\alignauthor
Luca Maria Aiello \\
       \affaddr{Yahoo Labs}\\
       \affaddr{Barcelona, Spain}\\
       \email{alucca@yahoo-inc.com}
}

\maketitle

\begin{abstract}
When providing directions to a place, web and mobile mapping services are all able to suggest the shortest route. The goal of this work is to automatically suggest routes that are not only  short but also emotionally pleasant. To  quantify the extent to which urban locations are pleasant, we use data from a crowd-sourcing platform that shows two street scenes in London (out of hundreds), and a user votes on which one looks more beautiful, quiet, and happy. We consider votes from more than 3.3K individuals and translate them into quantitative measures of location perceptions. We arrange those locations into a graph upon which we learn pleasant routes. Based on a quantitative validation, we find that, compared to the shortest routes, the recommended ones add just  a few extra walking minutes and are indeed perceived to be more beautiful, quiet, and happy. To test the generality of our approach, we consider Flickr metadata of more than 3.7M pictures in London and 1.3M in Boston, compute proxies for the crowdsourced beauty dimension (the one for which we have collected the most votes), and evaluate those proxies with 30 participants in London and 54 in Boston. These participants  have not only rated our recommendations but have also carefully motivated their choices, providing insights for future work.

% We are also able to compute a proxy for the beauty dimension from Flickr metadata associated with more than 3.7M pictures in London. 

% Then, to  quantify the extent to which locations are popular, we crawl all Foursquare venues in the entire city. 

%At times, however, when visiting a friend, we do not necessarily take the fastest route but might enjoy alternatives that, for example, offer beautiful urban sceneries. 

\end{abstract}




%\section{Introduction}\label{sec:01_intro}
\section{Introduction}\label{sec:01_intro}

At times,  we do not take the fastest route but enjoy alternatives that offer beautiful sceneries. When walking, we generally prefer tiny streets with trees over large avenues with cars. However, Web and mobile mapping services currently fail to offer that experience as they are able to recommend only shortest routes. 

To capture which routes people find interesting and enjoyable, researchers have started to analyze the digital traces left behind by users of  online services like Flickr or Foursquare. Previous work has, however, not considered the role of emotions in the urban context when recommending routes. Yet, there exists the concept of psychogeography, which dates back to 1955. This was defined as ``the study of the precise laws and specific effects of the geographical environment, consciously organized or not, on the emotions and behavior of individuals''~\cite{debord55}. The psychogeographer ``is able both to identify and to distill the varied ambiances of the urban environment. Emotional zones that cannot be determined simply by architectural or economic conditions must be determined by following the aimless stroll (\emph{derive})''~\cite{coverley2006psychogeography}. Mobile applications have been recently proposed to ease making \emph{derives} (i.e., detours in the city): these include Derive app\footnote{\tiny\url{http://deriveapp.com/s/v2/}}, Serendipitor\footnote{\tiny\url{http://serendipitor.net/site/}}, Drift\footnote{\tiny\url{http://www.brokencitylab.org/drift/}}, and Random GPS.\footnote{\tiny\url{http://nogovoyages.com/random_gps.html}} The goal of our work is to go beyond supporting \emph{derives} and to propose ways of automatically generating routes that are not only  short but also emotionally pleasant. This goal is not algorithmic but is experimental. We rely on crowdsourced measurements of people's emotional experience of the city and use those measurements to propose new ways of recommending urban routes. Despite emotional responses being subjective and difficult to quantify, urban studies have repeatedly shown that specific visual cues in the city context are consistently associated with the same fuzzy concept (e.g., with beauty)~\cite{de2008architecture,peterson67model,quercia14aesthetic,salesses13collaborative}. For example, previous work has found that green spaces and Victorian houses are mostly associated with beauty, while trash and broken windows with ugliness. 

To meet our research goal, we make three main contributions:
\begin{itemize}

%\item We build a  graph whose nodes are locations and whose edges connect geographic neighbors ($\S$\ref{sec:03_venuegraph}). These locations are then ranked based on whether they are emotional pleasant. To gather emotion scores, we build a crowd-sourcing platform that shows two street scenes in London (out of hundreds), and a user votes on which one looks more beautiful, quiet, and happy ($\S$\ref{sec:02_crowdsourcing}). We collect votes from more than 3.3K individuals and translate them into quantitative measures of urban perception (i.e., emotion scores).

\item We build a  graph whose nodes are locations and whose edges connect geographic neighbors ($\S$\ref{sec:03_venuegraph}). With this graph, we rank locations based on whether they are emotionally pleasant. The emotion scores come from a crowd-sourcing platform that shows two street scenes in London (out of hundreds), and a user votes on which one looks more beautiful, quiet, and happy ($\S$\ref{sec:02_crowdsourcing}). 

%We use votes from more than 3.3K individuals and translate them into quantitative measures of urban perception (i.e., emotion scores).

\item We quantitatively validate the extent to which our proposal recommends paths that are not only short but also emotionally-pleasing ($\S$\ref{sec:05evaluation}). We then qualitatively evaluate the recommendations by conducting a user study involving 30 participants in London. 

\item We finally test the generalizability of our proposal by: a) presenting a way of predicting the beauty scores from Flickr metadata; and b) testing the beauty-deri\-ved paths with our 30 participants in London and with a new group of 54 participants in Boston ($\S$\ref{sec:beyond}).

\end{itemize}

\section{Related work} \label{sec:01b_related}

\hyphenation{fa-mi-liar}
\hyphenation{back-tracking}
\hyphenation{recommen-ding}
\hyphenation{vi-si-ted}

Early research on route recommendation focused on finding the most  \emph{efficient} routes. For example, Chang \emph{et al.}~(\citeyear{Chang11discovering}) used a backtracking algorithm to recommend \emph{car} routes that deviate from a user's familiar/past trajectories. Ludwig \emph{et al.}~(\citeyear{Ludwig09Recommendation}) used an adaptive A*-like algorithm to recommend \emph{public transport} routes that afford both short walks and little waiting times. More recently, tools for recommending the safest or smoothest \emph{cycle} paths in the city have also been proposed~\cite{quaggiotto12bikedistrict}.

In addition to ways of recommending efficient paths, researchers have also investigated the problem of recommending \emph{distinctive} and \emph{interesting} urban routes~\cite{PeregrinoXXMApping,VanCanneyt11time}. The idea behind this line of work is to use geo-referenced online content (e.g., Flickr pictures) to learn and recommend popular trajectories~\cite{baraglia13learnext,Yoon10smart,Yoon12social}. De Choundry \emph{et al.}~(\citeyear{DeChoudhury10automatic}) and El Ali \emph{et al.}~(\citeyear{ElAli13photographer}) both used Flickr data to mine popular spatio-temporal sequences of picture uploads and to then recommend the corresponding urban routes. De Choundry \emph{et al.} identified the movements of individual tourists by tracking when and where they were uploading photos and by then using the resulting trajectories to connect points of interests in a graph. By embedding 
 location information such as average time spent at a location and location popularity, they were able to use an orienteering algorithm on the graph to compute the optimal  number of interesting locations to visit given a time budget. El Ali \emph{et al.} followed a similar idea: by clustering sequences of pictures uploaded at similar times, they used a sequence alignment algorithm borrowed from biology to produce trajectories containing interesting locations. 

In addition to using geo-located pictures, one could exploit GPS traces.  As opposed to social media, mobile phones enjoy high penetration rates and, as such, GPS traces can help  identify interesting places not only in cities but also in suburban regions~\cite{wei13exploring}. Zheng \emph{et al.}~(\citeyear{zheng09mining}) mined such traces by arranging both visited places and mobile users in  a bipartite graph, and then ranking places by graph centrality to extract top interesting regions. Their focus was on spotting interesting locations rather than routes. 

%However, they also suffer from slimited spatial resolution, as localizing users based on cell towers is far less precise than doing so with GPS-enable phones. 

All this line of work has been tailored to touristic use cases where paths can be considerably longer than the shortest ones~\cite{Schoning08evaluating}, and where  recommending frequently visited locations is a reasonable choice. 
%Moreover, given how these models are built, the recommended routes tend to stick to popular beaten paths and, so far, very limited research effort has gone into methodologies for evaluating route recommenders in the wild.

More recently, given the popularity of mobile social networking applications, researchers have been able to explore  personalization strategies for tourists and residents alike. Meng \emph{et al.}~(\citeyear{Meng12intention}) leveraged traces from Foursquare to plan itineraries that need to pass through different types of locations (e.g., restaurants, gas stations) and, given a user demand for some location types, they computed paths using an ant colony optimization algorithm. Cheng \emph{et al.}~(\citeyear{Cheng11personalized}) annotated historical data of traveled paths with demographic information and used a Baeysian learning model to generate personalized travel recommendations based on demographic segmentation. Kurashima \emph{et al.}~(\citeyear{Kurashima10travel}) addressed a somewhat similar problem - they profiled users according to their past travel histories. These approaches output sequences of locations according to different criteria but do not focus on the nature of the paths connecting those locations.

To date, there has not been any work that considers people's emotional perceptions of urban spaces when recommending routes to them. We thus set out to do such a work by collecting reliable perceptions of urban scenes, incorporating them into algorithmic solutions, and quantitatively and qualitatively evaluating those solutions.

\section{Our Proposal}
Our goal is to suggest users a short and pleasant path  between their current location $s$ and destination $d$. We meet this goal in four steps:
\begin{enumerate}
\item Build a graph whose nodes are all locations in the city under study ($\S$\ref{sec:03_venuegraph}).
\item Crowdsource people's perceptions of those locations along three dimensions: \emph{beautiful},  \emph{quiet}, and \emph{happy} ($\S$\ref{sec:02_crowdsourcing}).
\item Assign scores to locations along each of the three ($\S$\ref{sec:04_domainaware}).
\item Select the path between nodes $s$ and $d$ that strikes the right balance between being short and being pleasant ($\S$\ref{sec:bestpath}).
\end{enumerate}

\subsection{Building Location Graph} 
\label{sec:03_venuegraph}

%We divide the bounding box of London (up to travel zone 3) into 1,343 walkable cells, each of which is 200x200 meters in size.

We divide the bounding box of central London (travel zone 1\footnote{\tiny\url{http://visitorshop.tfl.gov.uk/help-centre/about-travel-zones.html}}) into 532 walkable cells, each of which is 200x200 meters in size. Previous research has established that 200m tends to be the threshold of walkable distance in urban areas~\cite{o1996walking,legible06}. In dense parts of London, such a distance would typically correspond to two blocks that could be covered by a 2.5-minute walk.  Having those cells at hand, we make them be nodes in  a location graph. Each node is a location and links  to its eight geographic neighbors.\footnote{\tiny Since a boundary cell would have less than eight neighbors, we  link it to  a number of  additional closest cells within the grid such that, as a result, it would link to eight nodes in total.} To quantify the extent to which a node reflects a pleasant location, we need to capture the way people perceive that  location, and we do so next.

\subsection{Crowdsourcing Perceptions} \label{sec:02_crowdsourcing}

We rely on the data gathered from a crowdsourcing web site to assess the extent to which different city's locations are perceived to be beautiful, quiet and  make people happy~\cite{quercia14aesthetic}. Available under \url{UrbanGems.org}, the site picks up two random urban scenes and ask users which one of the two is more beautiful, quiet, or happy. As for scenes, the site does not use Flickr images, as they considerably vary in quality, but taps into two image sources that offer  pictures of comparable quality: Google Street View pictures captured by camera-mounted cars, and Geograph\footnote{\tiny\url{http://www.geograph.org.uk/}} pictures provided by volunteers. To control for image bias, we perform two main steps. First, we make sure that multiple images from the two sources are available at \emph{each} location. Second, we check whether user ratings are not correlated with objective measures of image quality, and we indeed find that there is no correlation between images' ratings and two commonly used proxies for quality (i.e., sharpness and contrast levels~\cite{Yeh2010}).

At each game round, users should either click on one of the two scenes or opt for ``Can't Tell'', if undecided on which picture to click on. With each selection, the user is asked to guess the percentage of other people who shared their views, scoring points for correct guesses. Those points  can then be shared through the social media sites of Facebook and Twitter. To avoid the sparsity problem (too few answers per picture), a random scene is selected within a 300-meter radius from a subway station and within the bounding boxes of census areas. This results into 258 Google Street views and 310 Geograph images, all of which have ratings that are roughly normally distributed~\cite{quercia14aesthetic}. We use multiple images from the two sources at each location. By collecting a large number of responses across a large number of participants, we  are now able to determine which urban scenes are perceived in which ways along the three qualities.

The choice of the three qualities is motivated by their importance in the urban context according to previous studies. Being able to find \emph{quiet} places might ``promote `sonic health' in our cities and offer a public guide for those who crave a retreat from crowds''\footnote{\tiny \url{http://www.stereopublic.net/}}.
%, and this is important for those in need of quietness and for people with disabilities, like autism and schizophrenia.
As for \emph{beauty}, we are not  the first to measure its perceptions. In 1967, Peterson proposed a quantitative analysis of public perceptions of neighborhood visual appearance~\cite{peterson67model} and found that beauty and safety are approximately collinear. Finally, we choose happiness not least because urban studies in the 1960s tried to systematically relate well-being in the urban environment (i.e., happiness) to the fundamental desire for visual order, beauty, and aesthetics~\cite{lynch1960}. As a result, well-being or, more informally, happiness has taken centre stage in the scientific discourse for decades. 

%The platform was released publicly available and issued a press release in  September 2012. Shortly after that, the site was featured in major newspapers and news sites, including BBC News. 
The platform was released in September 2012 and after 4 months data from as many as 3,301 participants was collected: 36\% connecting from London (IP addresses), 35\% from  the rest of UK, and 29\%  outside UK. A fraction of those participants (515) specified their personal details: the percentage of male-female for those participants  is 66\%-34\%, the average age is 38.1 years old, and the racial segmentation reflects that of the 2001 UK census.\footnote{\tiny \url{http://en.wikipedia.org/wiki/Ethnic_groups_in_the_United_Kingdom}} Upon processing 17,261 rounds of annotation (each round requires to annotate at most ten pairs), we rank pictures by their scores for beauty, quiet, and happiness, and those scores are based on the fraction  of votes the pictures have received. The ranking is reliable because the number of annotators is $>$$3K$ and distribution of scores is normal with median as high as 171 for beauty, 12 for quiet, and 16 for happy. The number of answers for the three qualities is different as the default question is that on beauty, which thus preferentially attracts more answers.%, while the other two are accessible from a drop-down menu.
%Interestingly, despite that the happiness question comes last in the menu, it has still attracted more votes than the question on quietness, reflecting our users' preferences for happiness over quietness. 

We compute the correlations between each pairwise combinations of the three qualities.  All correlations are statistically significant (i.e., all $p$-values are $< 0.0001$) and are the following: happy-quiet has $r=0.29$, quiet-beauty $r=0.33$, and beauty-happy is $r=0.64$. As one would expect from the literature~\cite{de2008architecture}, we find that the strongest affiliation is that between beauty and happiness, so we should expect that the paths we will recommend for beauty and those for happiness might partly overlap at times.

%
%beauty = 0.37 + 0.03 log(density) + 0.20 log(posemo)/c - 0.21 log.negative/c
%
%where
%* density is the column density of flickr pictures in the file
%* posemo is the column posemo in the file (i think it's the 70th column in your file, but you need to double check - i attached an xls file that might help)
%* log.negative = log(swear) + log(negemo) + log(anx) + log(sad) + log(anger)
%* c=#classified words in LIWC (this is the sum of all words)

\subsection{Scoring Locations} \label{sec:04_domainaware}

To rank a location, we need to compute the likelihood that it will be visited because it is pleasant. One simple way of expressing that is with $p(go | happy) \propto     p(happiness | go) $.

%For each location, we compute   $p_{happy}$ with~(\ref{equation:p_happy}) . 
%
%\begin{eqnarray*} 
%& p &(go |  happy)  \propto \\
%& \propto &  p_{happy}  \\
% & \propto &   k_2    \cdot h_i^3 
% \end{eqnarray*}

 The probability $p(happiness | go)$ captures the idea that individuals visits locations that make them happy. We thus need a way to measure a location's happiness and, to that end, we resort to our crowdsourced  scores ($\S$\ref{sec:02_crowdsourcing}). More specifically, given the crowdsourced happiness score $h_i$ for cell $i$, we compute the corresponding happiness probability with a curve that is, for example, cubic:
% Figure~\ref{fig:answers-perscene_beauty} shows the distribution of those scores, 
\begin{equation}
%p_{happiness}= k_2 \cdot h_i^3, 
p(happiness | go) = k \cdot h_i^3, \textrm{ where }  k=\frac{1}{ max\{  h_i^3 \} \forall_i }
\label{equation:p_happy}
\end{equation}
Thus, the higher a location's crowdsourced score, the happier it is likely to be. 
By substituting $h_i^3$ with any of the expressions in Table~\ref{tab:curves}, we obtain the alternative happiness scoring functions.  We apply  those scoring functions  to the two remaining scores of quiet and beauty too. We do so by simply substituting $h_i$ with $q_i$ (location $i$'s quietness score) and with $b_i$ ($i$'s beauty score).  Next,  for brevity, we will report only the results for the cubic curve for which our experiments have shown the highest percentage improvements.

\begin{table}[!tb]
\begin{center}
\begin{tabular}{|c|c|}
\hline 
\textbf{Name}  & \textbf{Formula} \\
\hline 
Linear & $h_i$ \\
\hline 
Cubic & $h_i^3$ \\
\hline 
Exponential & $e^{h_i}$ \\
\hline 
Square root & $\sqrt{h_i}$ \\
\hline 
Sigmoid & $\frac{1}{1 + e^{-h_i}}$ \\
\hline 
\end{tabular}
\caption{Five expressions experimentally used to map crowdsourced scores to probabilities. With a location's crowdsourced scores of happiness, those expressions return the location's likelihoods of being considered happy if one were to visit it. } \label{tab:curves}
\end{center}
\end{table}

%\begin{table}[!tb]
%\begin{center}
%\begin{tabular}{c|ccccc}
 %\textbf{Path type}  & \textbf{$\Delta$Happy}  &  \textbf{$\Delta$Beauty} & \textbf{$\Delta$Quiet}  & \textbf{$\Delta$Dist} & \textbf{Iter} \\
%\hline 
%Happy & 0.296 & 0.237 & 0.024 & 0.134 & 1500 \\    
%Beauty & 0.223 & 0.300 & 0.056 & 0.132 & 1250 \\
%Quiet  & 0.047 & 0.095 & 0.259 & 0.141 & 1900 \\
%Flickr & - & 0.111 & - & - & - \\
%\end{tabular}
%\caption{Results over 190 paths with the selected stopping condition. Values express relative icrease over the shortest path values.} \label{table:results}
%\end{center}
%\end{table}

%Also, we apply the same scoring functions  to the beauty scores ($beauty^\prime$) predicted  from Flickr features, as per expression~(\ref{equ:linear_beauty}). By doing so, we will test the extent to which our proposal could be extended to cities other than London for which no crowdsourcing game but Flickr metadata is available. 

\subsection{Selecting Best Path} \label{sec:bestpath}

Upon the location graph and having the likelihood of visiting each location, we now select the best path from source $s$ to destination $d$ in four steps:

\emph{Step 1.} Identify $M$ shortest paths between $s$ and $d$. To identify them, we run Eppstein's algorithm~\cite{Eppstein99} and find the $M$ shortest paths connecting each pair of nodes $s$ and $d$. To be sufficiently ``exhaustive'', we initially set $M$ to be as high as $10^6$. This choice makes it possible to explore the full set of solutions, including any of the solution that alternative approaches might return (e.g., orienteering algorithms). 

\begin{figure}[!t]
 \includegraphics[width=\columnwidth]{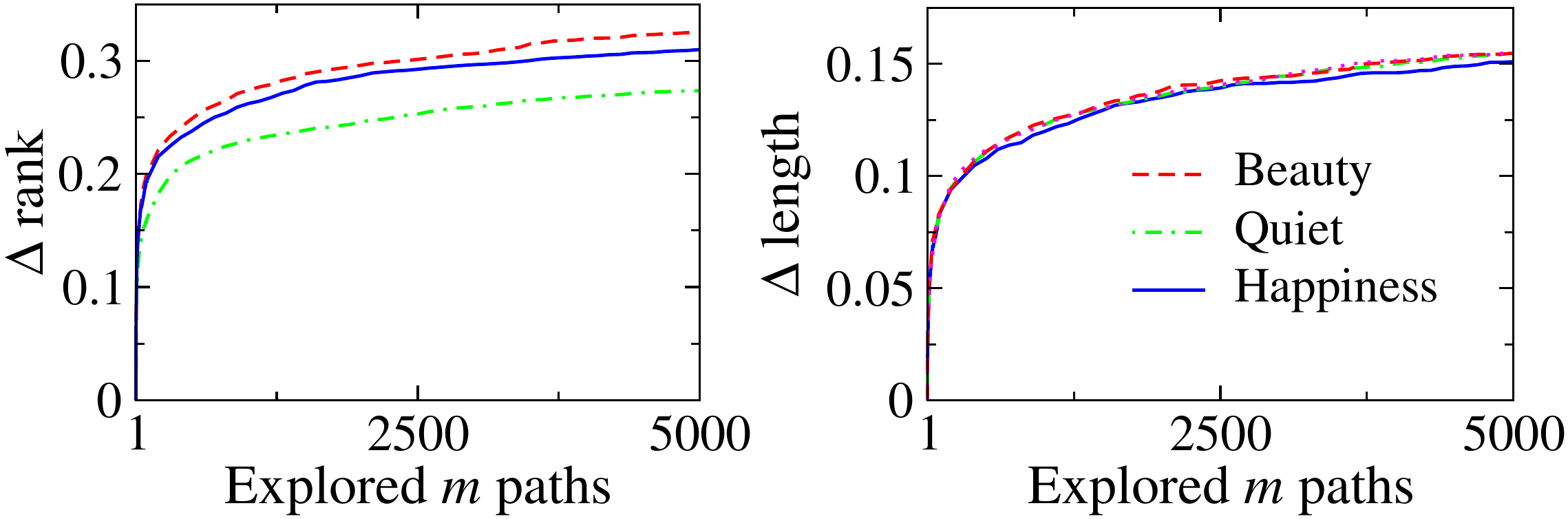}
 \caption{$\Delta$rank and $\Delta$length for different exploration levels $m$ (left panel). Respectively, these represent improvements in beauty/quiet/happiness ranks (left panel) and demand for extra walking time if only the first $m$ shortest paths are considered (right).}
 \label{fig:iter_vs_all}
\end{figure}

\emph{Step 2.}  Compute the \emph{average} rank for all locations in each of the first $m$ paths (with $m \le M$). For computational tractability, we do not consider all the $M$ paths at once but iteratively explore  the first $m$ paths. At each exploration,  we record the path with the lowest (best) average rank. Such a path exploration has, of course, diminishing returns (Figure~\ref{fig:iter_vs_all}, left panel): the more paths we consider (the higher $m$), the less likely the best value for rank will change. This suggests that it is not necessary to explore all $M$ paths but we can explore a tiny subset of them without loss of performance, and that is what the next step does. 
 
\emph{Step 3.}   Terminate when the average rank improves less than $\epsilon$. To select $\epsilon$, we use The Marginal Value Theorem (MVT). This is an analytical tool for optimizing the trade-off between benefit (rank improvement $\Delta$rank) and cost (exploration of the first $m$ paths). One can show analytically that, for the function of rank vs. $m$, it is best to keep increasing $m$ only until $\frac{\Delta \textrm{rank}}{\Delta m}$ equals $\frac{ \textrm{rank}}{m}$; after that, one should terminate and take the path among those considered that has the best average rank. 

\emph{Step 4.}  Select the path that has so far been found to have the best rank.

By repeating these steps for each of the three ranks (beauty, quiet, and happiness), we obtain three paths between $s$ and $d$ in addition to the shortest one (to the baseline).
%
%Since the average rank is expressed for three qualities, in addition to the shortest path, we obtain 
%
%
%for each  $s$ and $d$

\begin{figure*}[htb] \begin{center}
\subfigure[Shortest]{\includegraphics[width=.48\textwidth]{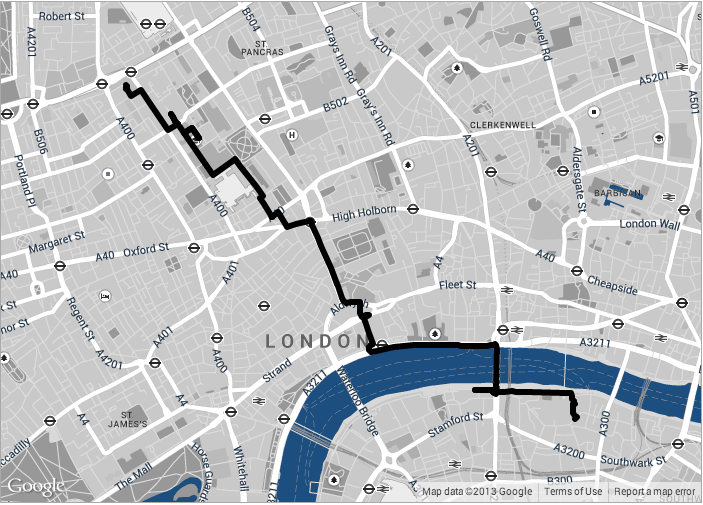} \label{fig:map_shortest} }
\subfigure[Beauty]{\includegraphics[width=.48\textwidth]{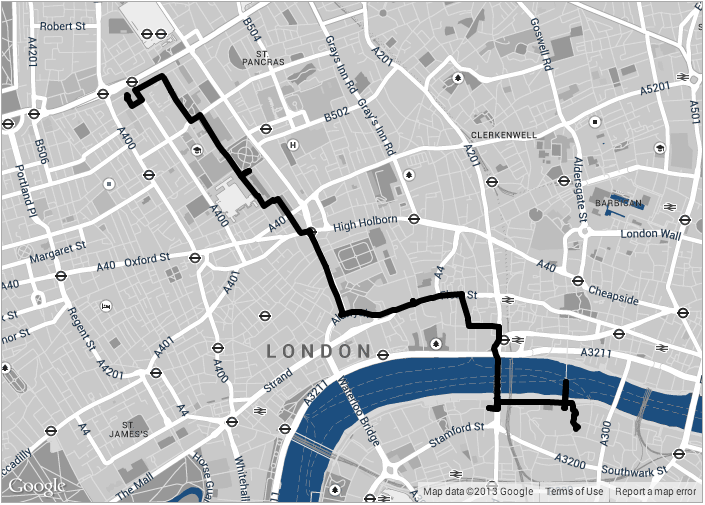} \label{fig:beauty_path} }
%\subfigure[Beauty (Flickr)]{\includegraphics[width=.23\textwidth]{figs/beauty_flickr_path} \label{fig:beauty_flickr_path} }
\subfigure[Quiet]{\includegraphics[width=.48\textwidth]{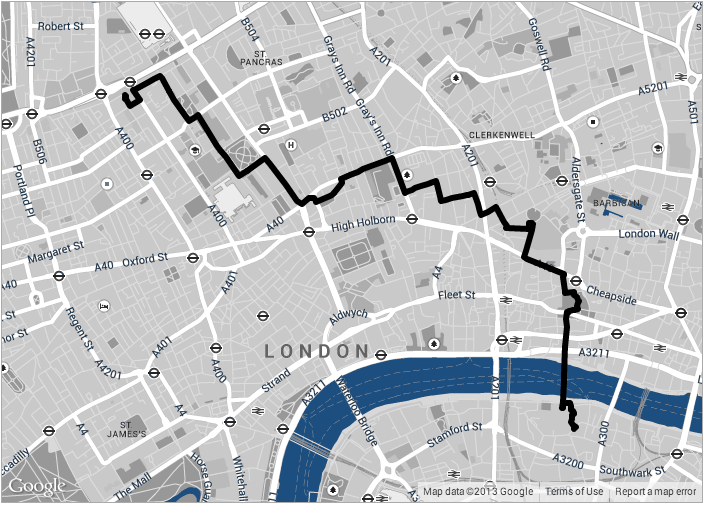} \label{fig:quiet_path} }
\subfigure[Happy]{\includegraphics[width=.48\textwidth]{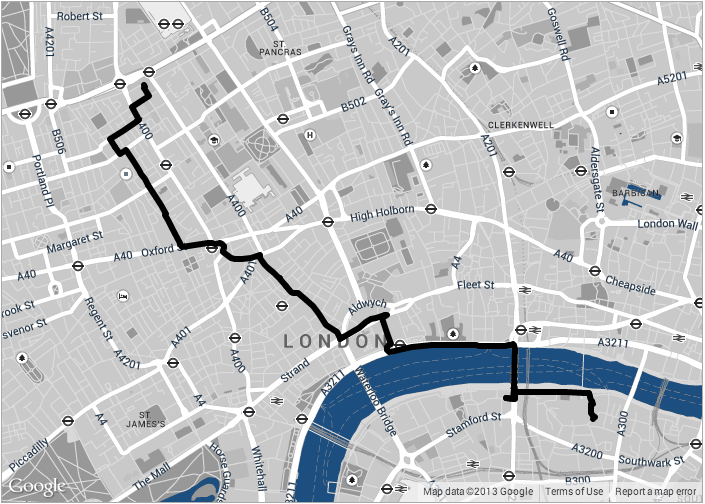} \label{fig:happy_path} }
\end{center} 
\caption{Maps showing the different paths between Euston Square and Tate Modern.}
\label{fig:summary_paths} 

\end{figure*}

\section{Evaluation} 
\label{sec:05evaluation}
%======================

The goal of our proposal is to recommend paths that are not only short but also pleasant. To ascertain the effectiveness of our proposal at meeting this goal, our evaluation ought to answer three main questions: \\
\emph{(Validation)} Is our proposal able to recommend paths that are pleasant? ($\S$\ref{sec:accuracy}) \\
\emph{(Length trade-off)} Are pleasant paths considerably longer than shortest ones? ($\S$\ref{sec:trade-off}) \\
\emph{(User assessment)} Do people find the recommended paths to  be actually  beautiful, quiet, and happy? ($\S$\ref{sec:user_assessment})
 
\subsection{Validation}
\label{sec:accuracy}

\begin{table}[!tb]
\begin{center}
\begin{tabular}{l|ccc}
 \textbf{Path recommended for}  & \textbf{$\Delta$Beauty}  &  \textbf{$\Delta$Quiet} & \textbf{$\Delta$Happy}   \\
\hline 
Beauty & \textbf{30\%} & 6\%  & 22\%   \\
Quiet  &  10\% & \textbf{26\%}  & 5\% \\
Happy  & 24\% & 2\% & \textbf{30\%} \\    
\end{tabular}
\caption{Percentage improvements for beauty, quiet, and happiness in the recommended paths  over the shortest ones  (results averaged across 190 paths).} \label{table:results}
\end{center}
\end{table}

We compare how  our recommended paths differ from the shortest ones in terms of average beauty, quiet and happy scores. We choose shortest path as baseline since it reproduces existing approaches that focus on time efficiency. To make this comparison computationally tractable, we consider twenty nodes that correspond to popular landmarks and cover the entire central part of the city under study.
We analyze the 190 paths resulting  from each node pair. As one expects, the paths recommended for beauty, quietness, and happiness all show  a percentage increase in the corresponding dimension (Table~\ref{table:results}). More specifically, compared to the shortest paths, beautiful ones are, on average, 30\% more beautiful  (and are happier as well); quiet paths are 26\% quieter; and happy paths are, again,  30\% happier  (and are  also more beautiful).  From these results, we conclude that not only our proposal effectively biases paths in the way we expect (which should be the case for any working solution) but also it does so to a considerable extent: with an increase that goes from $26\%$ to $30\%$. Now the question is whether this bias comes at the price of considerable longer paths. 

%After answering that question next, we shall also administer a survey to test whether individuals actually perceive those  increases  for average beauty, quiet and happy.

%\begin{table}[!tb]
%\begin{center}
%\begin{tabular}{c|ccccc}
% \textbf{Path type}  & \textbf{$\Delta$Happy}  &  \textbf{$\Delta$Beauty} & \textbf{$\Delta$Quiet}  & \textbf{$\Delta$Dist} & \textbf{Iter} \\
%\hline 
%Happy & 0.266 & 0.209 & 0.019 & 0.120 & 1500 \\    
%Beauty & 0.192 & 0.266 & 0.050 & 0.120 & 1250 \\
%Quiet  & 0.035 & 0.080 & 0.223 & 0.120 & 1900 \\
%Flickr & - & - & - & - & - \\
%\end{tabular}
%\caption{Results over 190 paths with the selected stopping condition. Values express relative icrease over the shortest path values.} \label{table:results}
%\end{center}
%\end{table}

%width=0.48\textwidth
%width=\columnwidth

\begin{figure}[!t]
\center
 \includegraphics[width=.75\columnwidth]{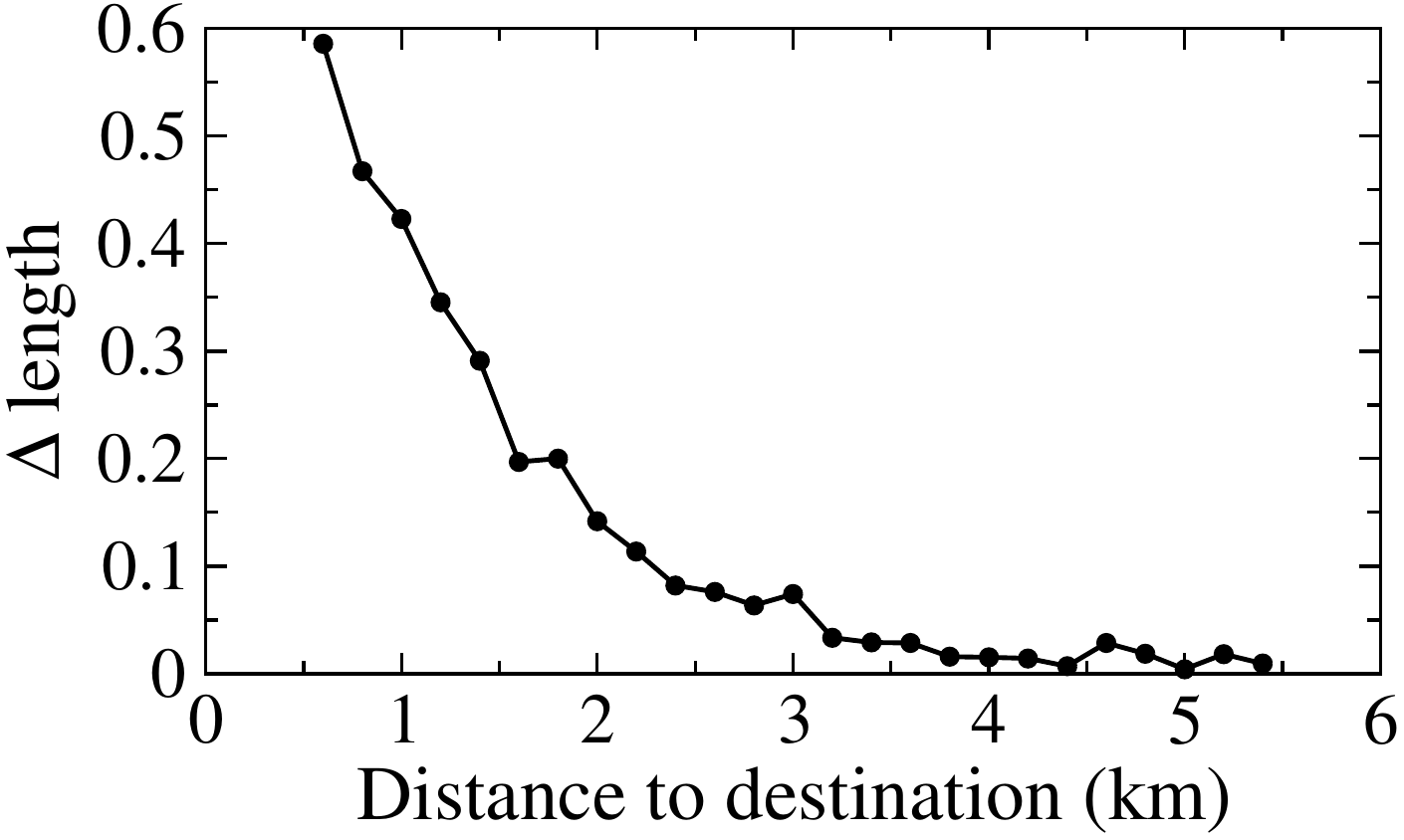}

 \caption{Additional distance (\% over shortest paths) required to recommend pleasant paths for destinations at different distances ($km$). As expected, for closer destinations, the additional cost  is larger than that for farthest destinations. }
 \label{fig:delta_distance}
\end{figure}

\subsection{Length Trade-off}
\label{sec:trade-off}

We test whether the recommended paths are considerably longer than the shortest ones. We find that, on average, the recommended paths are only 12\% longer. This is a good result for two main reasons. First, it is far lower than what  previous work reported, which admittedly focused on different contexts such as tourism or entertainment. For example, paths recommending touristic attractions tended to be half-a-day touristic experiences (twelve hours)~\cite{DeChoudhury10automatic}, and those capturing people's salient experiences tended to be  60\% longer than the shortest paths~\cite{ElAli13photographer}. Second, the increase of 12\% in length practically translates into  about 2 and a half additional cells, which correspond to roughly 7 and a half additional minutes. These numbers are all average values and, as such, they collate  both long and short paths together. To break down those results,  we plot the extra length required for recommending pleasant paths (over shortest)  for destinations at increasing distances (Figure~\ref{fig:delta_distance}).  We expect that the  farther the destination, the lower the extra time required by our recommendation over the shortest paths. We find that, for  1km ($x$-axis), paths tend to be 40\% longer ($y$-axis), and, more importantly, that extra cost indeed decreases exponentially. That is because, to identify pleasant sceneries, the shorter a path, the more deviations (and extra walking time) are required; by contrast, in a longer path, there is plenty of room for finding pleasant sceneries without imposing any additional walking overhead.

%\begin{figure}[!t]
% \includegraphics[width=\columnwidth]{./figs/shortest_path}
% \footnotesize 
% \caption{Map showing the path between Euston Square and Tate Modern.}
% \label{fig:map_shortest}
%\end{figure}

\begin{figure}[t]
 \includegraphics[width=\columnwidth]{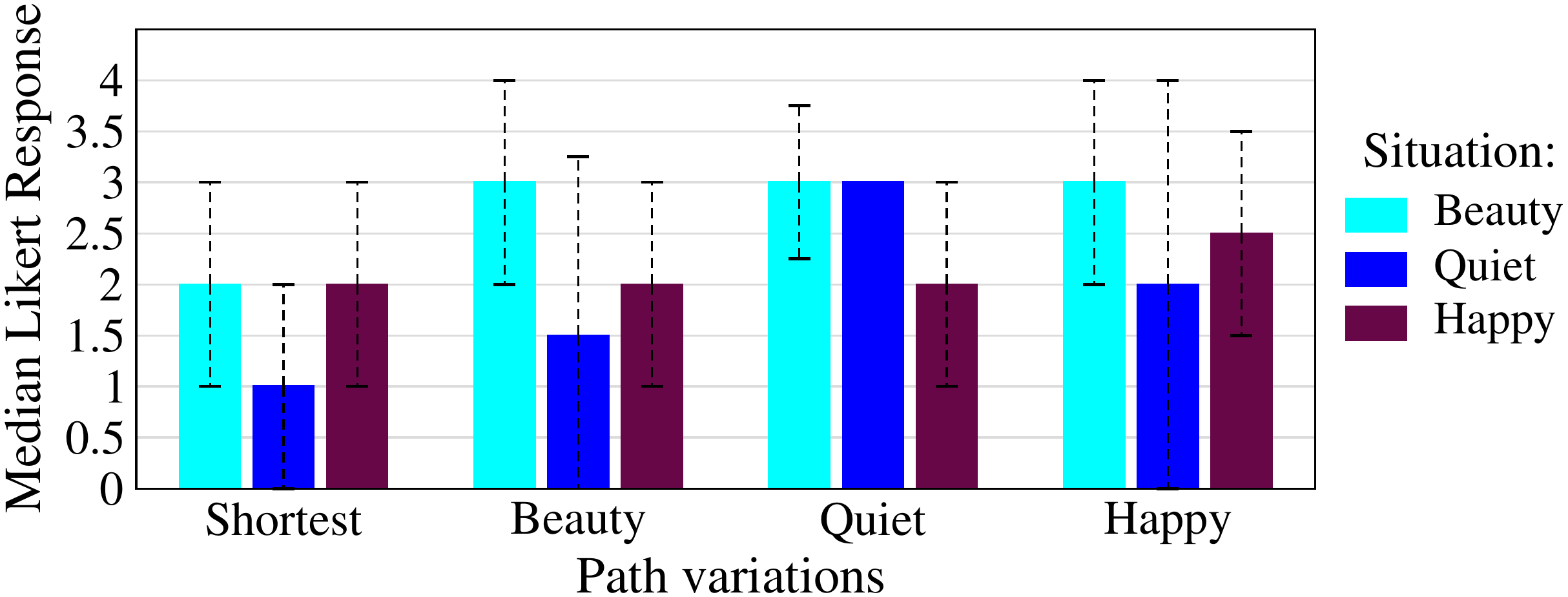}
 \footnotesize 
 \caption{Median of Likert responses for four path variations (shortest, beauty, quiet, and happy) in the three scenarios. Error bars reflect the range across quartile, which measures dispersion. All results are statistically significant. }
 \label{fig:bars_survey}
\end{figure}
% In each of the four variations (e.g., quiet variation), the dispersion associated with the intended quality (e.g., error on the blue bar reflecting quiet) is lower than the dispersions of the other two qualities.

%======================
\subsection{User Assessment}
\label{sec:user_assessment}
To evaluate whether our recommendations are perceived by individuals as desirable alternatives to current shortest route planners, we resort to a mixed-method user study in which both quantitative and qualitative measures are extracted. The goal of the study is to test four paths -  shortest, beautiful, quiet, and happy - between the same source and destination. We have to keep the two end points fixed to avoid rating sparsity.

\newpage
\subsubsection{Experimental Setup and Execution}
\label{sec:setup}

Our participants will see the four paths, like the ones shown in Figure~\ref{fig:summary_paths}, on a web page but do not know which one is what. The two end points of  the paths are Euston Square and Tate Modern.
We  chose those two specific points  because the resulting paths: 1) are between two locations well-known to Londoners;  2) are in central London, and that increases the chance participants will know them; 3) are at walking distance; and 4) go across the River Thames, allowing us to test any potential effect of the mental divide between North and South London~\cite{Quercia13psycho}.

To begin with, we ask each of our participants to read a consent form and optionally provide  age, gender, years living in London, and email address. They are also asked to tell us the extent to which they are familiar with the five paths using a Likert scale (i.e., they have to  choose among Strongly disagree, Disagree, Neither agree nor disagree, Agree, and Strongly agree). After this initial step, we test the four routes in three different scenarios, each of which corresponds to one of the three fuzzy qualities -  beauty, quietness, and happiness - and is meant to emulate a realistic use of a navigation service. For the \emph{happy scenario}, a participant has to imagine to be in the company of a friend who is a bit down. Given that, the participant  decides to bring the friend to the Tate Gallery and has to assess which of the five paths would make this friend happier using, again, a Likert scale. For the \emph{quiet scenario}, the participant wishes to go to the Tate. Today (s)he has been exposed to many people and lots of car traffic in central London, and thus (s)he wishes to find a quiet path to get there. The participants has to assess the extent to which each  of the five paths is expected to be quiet on a Likert scale. Finally, for the \emph{beauty scenario}, the participant has to imagine that his/her best friend is visiting from Italy, and this friend loves beautiful things. The participant has to say which of the five paths is expected to offer beautiful views.  After answering each of the 5x3 questions, our participants can  motivate each of their answers under a free-text box.
% titled ``Could you say why you think so?''.

\subsubsection{Demographics of Participants}
Among our 30 participants in London, the percentage of male-female is 58\%-42\%. The most common age band is that of 30-35, which includes half of our participants; by contrast, only 5\% are below 23 years old, and 10\% above 48. As for familiarity with London, our respondents have lived in London, on average, for two and a half years.  Since this sample of users relates only to specific social groups,  it would be interesting to compare their demographics to those of social media users. From the latest Ignite report on social media~\cite{ignite12socialnet},  Foursquare and Twitter users tend to be university educated 25-34 year old women (66\%  women for Foursquare and 61\% for Twitter), while  Flickr users tend to be university educated 35-44 year old women (54\%  women). The demographics of our respondents thus show a skew towards 30-35 year old men (58\%  men). As such, compared to social media users, our respondents reflect a slightly older age group and are more likely to be men. 

%The demographics of subway passengers is by far the most representative of the London general population, being slightly skewed towards male with above-average income in the two age groups of 25-44 and 45-59~\cite{tfl}. 

\subsubsection{Quantitative and Qualitative Results}

As typically done, we  treat  our Likert responses as ordinal data and collate them into bar charts. We summarize the central tendency by the median and the dispersion  by the range across quartiles (error bars), and show the results in Figure~\ref{fig:bars_survey}. We will present the results that are statistically significant. We find that across the three scenarios, the shortest path performs worst, with median from 1.5 to 2 and dispersion $ \pm 1$. The best performing variation across the three scenarios is the happy variation with median as high as 3.  Finally, the clear-cut scenario is that of quiet, in which the quiet variation is indeed the one most preferred  by our respondents whose ratings have, in this case,  no dispersion: for the quiet path, we find a median score $3 \pm 0$ as opposed to  
$1 \pm 1$ for the shortest path, $1 \pm 1.75$ for the beauty variation, and  $3 \pm 2$ for the happy variation. These results suggest that, without telling which path is what, our participants readily associate the path to the intended quality of quiet, beauty, or happiness. 

\begin{figure}[!t]
\center
 \includegraphics[width=.99\columnwidth]{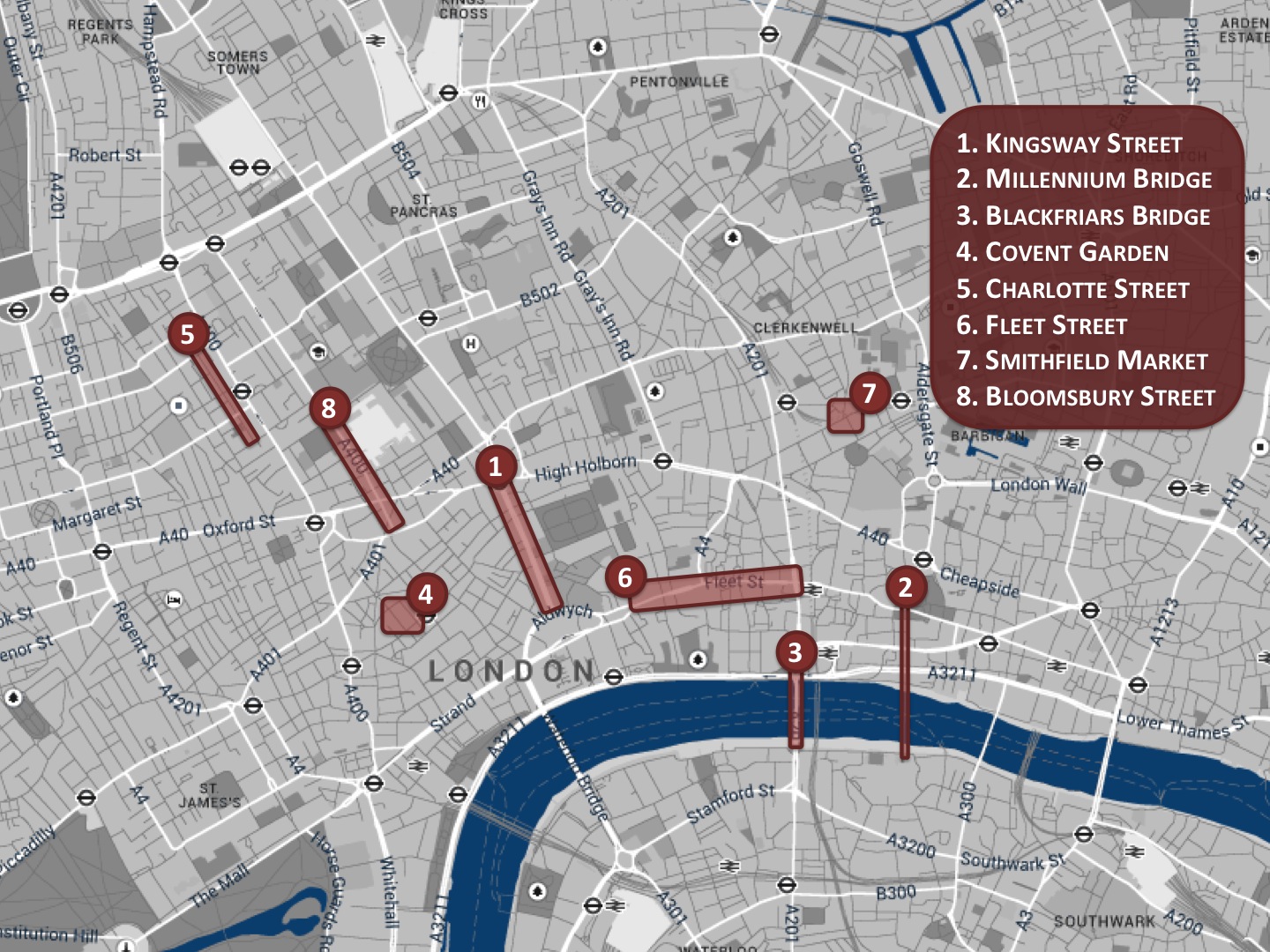}
 \footnotesize 
 \caption{Map of London with eight frequently mentioned places.}
 \label{fig:london_map_cut}
\end{figure}

To gain further insights into their choices, we have asked the participants to motivate their ratings (the map in Figure~\ref{fig:london_map_cut} marks  the places they frequently mentioned). All the respondents do not perceive any extra walking time for the \break quiet/beautiful/happy paths. Both beautiful and happy paths are associated with \emph{peaceful} places (e.g., back streets around St Paul's cathedral, the walkable Millenium Bridge as opposed to the car-infested Blackfriars Bridge). By contrast,  \emph{busy} places make people unhappy. By busy, our participants mean streets full of cars. For example, they consider the shortest path unpleasant because it was going \break through Kingsway, which is a heavily trafficked road in central London. However, at times, by busy, they also mean a street full of people. One participant considers a path undesirable because it goes through Kingsway and Fleet street: ``Kingsway is always busy with cars, and  Fleet street with pedestrians''. Also, beautiful and happy paths are positively associated with  \emph{historical} places (e.g., the Bloomsbury area). Interestingly, places that mix the two contrasting qualities of being \emph{historical yet busy} lead to controversial feedbacks. Most of our respondents consider Covent Garden and Charlotte street in extremely positive ways:  ``There is certainly colour in Charlotte street and some beautiful shops and events happening though Covent Garden''; ``Charlotte street is funky'';  ``I am very fond of Charlotte Street and Endell Street: they contain exciting and charming cafes and shops''. Yet, at times, the same very street acquires negative connotations: ``I would not think a busy Charlotte street would make the friend happier''. Overall, we note that  the presence of charming shops and historical elements balances the otherwise negative perceptions of busy places. Indeed the respondent who thought that Charlotte street would not make the friend happier adds: ``\ldots However the presence of people on a street packed with businesses might contribute to happiness''. Also, places offering distinctive experiences are considered happy. So is considered Soho: it ``offers a different view of London''; another respondent adds ``I like Soho and the area around Covent Garden. I think that walking in these areas would make my friend happier''. The concept of happiness also captures hard-to-quantify contextual aspects of urban life such as smell. As for unhappy places recommended by the shortest path, one respondent recalls that ``Southampton row is smelly'' and should not be recommended.
%The most recurring distinctive place was the River Thames: ``The picturesque stroll along the river bank would help cheer up a bit''.
Our respondents also point out that the experiences of a place might change during the course of a day: ``Fleet street is beautiful because of its history. However, depending on the time of day, it can be colourless and busy leading to the opposite results''. Finally, the urban quality among the three that results in the least controversial feedback is that of quiet as it is readily associated with parks and back roads. To sum up, a desirable place tends to be  peaceful, historical, and distinctive, whilst avoiding being too busy. Interestingly, places that either mix contrasting qualities (e.g., historical/charming \emph{vs.} busy) or experience drastic changes over time (e.g., busy during the week, and lovely in the weekend) yield mixed results.

\section{Crowdsourcing and London} \label{sec:beyond}

\begin{figure*}[t!] \begin{center}
\subfigure[Flickr Beauty in London]{\includegraphics[width=.32\textwidth]{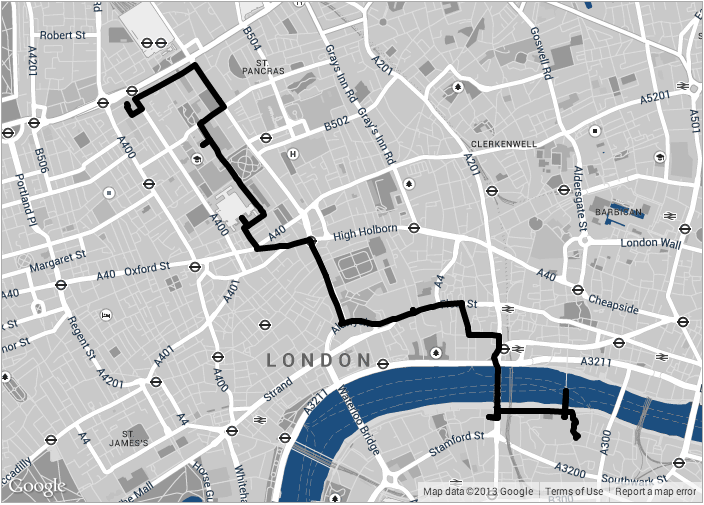} \label{fig:beauty_flickr_path_london} }
\subfigure[Flickr Beauty in Boston]{\includegraphics[width=.32\textwidth]{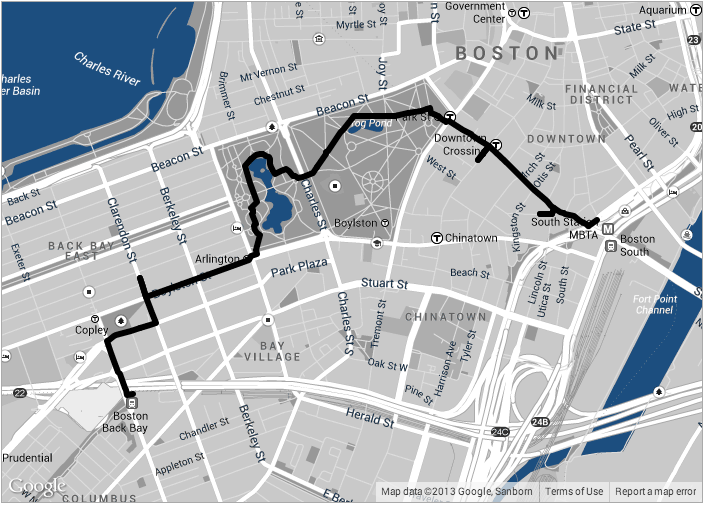} \label{fig:beauty_flickr_path_boston} }
\subfigure[Shortest in Boston]{\includegraphics[width=.32\textwidth]{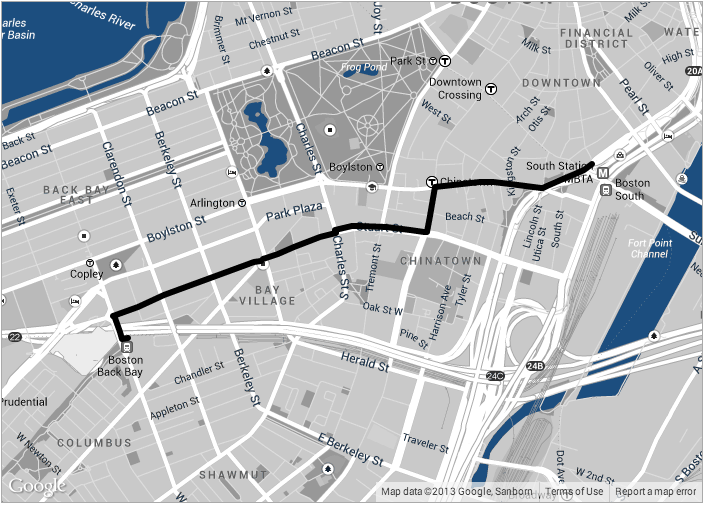} \label{fig:shortest_path_boston} }
\end{center} \vspace*{-6mm}
\caption{Maps showing the different paths with beauty scores predicted from Flickr's metadata.}
\label{fig:summary_flickr_paths} 
\vspace*{-1mm}
\end{figure*}

Critics might rightly claim that this work suffers from two main drawbacks. First, it relies on crowsourced ratings whose collection requires intense user involvement. Second, our evaluation has so far focused on a single city. To partly address those two concerns, in this section, we present  a way of predicting the beauty scores from Flickr metadata ($\S$\ref{sec:beyond_flickr}) and test these predictions not only in London but also in the city of Boston ($\S$\ref{sec:beyond_boston}).

\subsection{Beyond Crowdsourcing: Flickr}
\label{sec:beyond_flickr}

We test whether one can  predict beauty scores out of Flickr metadata. We choose beauty over quiet or over happiness since, being the default question, it has received the highest number of votes. We gather a random sample of 7M geo-referenced Flickr pictures within our bounding box of London, 5.1M of which have at least one tag and, as we shall see, 3.7M can be used for our purposes. For each of our locations (cells), we gather these statistics: number of pictures (density), number of views, of favorites,  of comments, and of tags received by those pictures.  We also get hold of the actual tags, clean them (i.e., convert them to lowercase and  remove  those  that are stop-words), and process them using a dictionary called ``Linguistic Inquiry Word Count''~\cite{pennebaker2013secret}. LIWC is a standard dictionary of 2,300 English words that capture 80\% of the words used in everyday conversations and reflect people's emotional and cognitive perceptions. These words fall into 72 categories, such as positive and negative emotional words, and words about work, school, money. Rather than grouping  words based on their material subject matter (e.g., `sports', `technology'), LIWC categories are generally abstract, and are based on linguistic and psychological processes. For example, there exist categories for cognitive processes (such as `insight' and `certainty'), psychological constructs (e.g., affect, cognition), as well as personal concerns (e.g., work, home, leisure activities). Each word may thus belong to multiple categories; for example, the prefix entry `hostil*' belongs to the categories `affect' (affective processes), `negemo' (negative emotions) and `anger'.  Having this dictionary at hand, we  count the number of  tags matching the 72 categories, and, after doing so, we are left with 3.7M pictures that have at least one classified tag. For each location (cell), we compute the \emph{normalized count} for each LIWC category $c$:
\begin{eqnarray} 
f_c= \frac{w_c - \mu_c}{\sigma_c} 
\label{equ:norm_count}
\end{eqnarray} 
where $w_c$ is the fraction of tags classified in category $c$ (over the total number of classified tags) for the location; $\mu_c$ is the fraction of tags in category $c$, averaged across all locations; and $\sigma_c$ is the corresponding standard deviation.  

Out of the Flickr features and LIWC categories, those that are significantly correlated with beauty scores are density (number of pictures in a cell), `posemo', `negemo',  `swear', `anx' (anxiety), `sad', and `anger'. We thus use the presence of those categories as predictors in a linear regression whose dependent variable is the beauty score. We find the following coefficients:
\begin{eqnarray} 
beauty^\prime = 0.37 + 0.03 \cdot log(density) + 0.20 \cdot f_{p} - 0.21 \cdot f_{n}
\label{equ:linear_beauty}
\end{eqnarray} 
where $ f_{p}$ is the fraction of tags containing positive emotions, and $f_n$ is the fraction of tags containing emotions in the categories `negemo',  `swear', `anx', `sad', and `anger'.  We should concede that  statistical models more sophisticated than a linear regression could have been used: for example, one might well decide to exploit the spatial autocorrelations of the beauty scores. However, given that  our goal is to carry out a preliminary study on Flickr, we opt for a linear regression, not least because it yields predictions that are easy to explain (as opposed to black-box machine learning approaches) and that are reasonably accurate. The $R^2$ of the regression is indeed $0.31$, which means that as much as 31\% of the variability of the beauty score can be solely explained by the presence of Flickr tags. 

\begin{table}[!tb]
\begin{center}
\begin{tabular}{|p{1.4cm} |p{2.8cm} | p{3.2cm}|}
\hline
 \textbf{Category}  & \textbf{Example words}  &   \textbf{Description}    \\
\hline 
posemo  &  happy, pretty, good &  Express pos. emotions\\    
\hline 
negemo   & hate, worthless  &  Express neg. emotions\\
\hline 
swear   & c**t, f**k  &  Swear words \\
\hline 
anx   & afraid, scare  &  Express anxiety \\
\hline 
sad   & cried, deprived  &  Express sadness\\
\hline 
anger   & abusive, disgusting  &  Express anger\\
\hline
\end{tabular}
\caption{Categories that reflect the use of language and whose presence in Flickr tags correlates with the beauty scores. } \label{table:categories}
\end{center}
\end{table}

\subsubsection{Results}
To build paths from Flickr metadata,  we use the Flickr-derived beauty scores and run the steps  described in Section~\ref{sec:bestpath}. We find that the resulting beautiful paths are, on average, 28\% more beautiful than the shortest ones. To see how users perceive a Flickr-derived path, during our user study in London (Section~\ref{sec:user_assessment}), we  had added a fifth path (in addition to the four we have already presented): a Flickr-derived beautiful path between Euston Square and Tate Modern (Figure~\ref{fig:beauty_flickr_path_london}). Our 30 respondents rate this path with neutral answers (median 2) but little uncertainty (0 dispersion). That is, our respondents do not feel strongly about the recommendation based on Flickr but  agree with each other when judging it, suggesting that Flickr-derived recommendations can be used as general-purpose suggestions when personalization is not possible (e.g., in cold-start situations). However, by comparing those who rated the Flickr variation above the median and those who rated it below, we find that these two classes of individuals differ. The Flickr variation is preferred by:
\begin{description}
\item[Women.] The percentage of male-female is 25\%-75\% for above-median raters, and 60\%-40\% for those below;
\item[Short-term residents.]  The average number of years living in London is 2 and a half for those above, and 3 and a half for those below. 
\end{description}
%That is, women and those who have lived in London, on average, one year or less tend to prefer the Flickr variation. 

% 1.6% of london population uses flickr; 4.2% in boston
\mbox{ }
\subsection{Beyond London: Boston}
\label{sec:beyond_boston}

London is in itself quite a peculiar city: from a geo-demographic perspective, London is by far the largest city in the EU (the most densely populated within city limits), with a diverse range of people and cultures spread all over its area. To complement our London study, we present in this section a brief study in the city of Boston, chosen for their very different characteristics. From a geo-demographic perspective, we scale both population density and area size of a factor down  with respect to London. From a Flickr perspective,  picture density is slightly higher in Boston (1.1 more pictures per $m^2$ than in London), while engagement with content is slightly lower (a London picture has, on average,  0.8 more tags than one in Boston).

% 1.1 more pictures than in London
% 0.8 more tags per picture than in Boston

% (5.6 \emph{vs.} 4.5 in London)
% (e.g., 5.4 tags per picture vs. 4.6 in Boston).

\subsubsection{Quantitative Results}
By running the steps  in $\S$\ref{sec:bestpath} on the metadata in Boston, we find that the Flickr-derived beautiful paths are, on average, 35\% more beautiful than the shortest paths. This is higher than any improvement we have experienced for any of the three qualities in London, suggesting that routing through beautiful sceneries (and, accordingly, avoiding ugly ones) is easier in Boston than it was in London. Indeed, American cities (including Boston) tend to offer urban situations far more diverse in terms of safety and beauty  than what European cities do~\cite{salesses13collaborative}, and, for the routing algorithm, diversity translates into more options to choose from. 

In Boston, we repeated  the London's user study ($\S$\ref{sec:setup}): we were able to recruit as many as 54 participants who compared a Flickr-derived beautiful path (Figure~\ref{fig:beauty_flickr_path_boston}) with the shortest one (Figure~\ref{fig:shortest_path_boston}) using, again, a Likert scale. These two paths go from Back Bay Station to South Station,  and the Flickr-derived is chosen to be the one that has the highest improvement in average beauty. 

%male=31, female=23
% 24-29= 17 (32%)
% 30-35=23 (43%)
% 36-41= 4
%42-47=6
%48+ = 3
% total age=53 (one did not say)

The percentage of male-female participants overall is 57\%-43\%. Also, the most common age band is that of 30-35, which includes 43\% our participants; by contrast,  32\% are below 30 years old, and only 17\% above 42. As for familiarity with Boston, our respondents have lived in the city, on average, for four and half years.

From our respondents' ratings, we find that, again, the shortest path performs worst, with a median of 1 and dispersion $ \pm 1$. Thus the best performing is the Flickr-defined path with median as high as 3 $ \pm 1$, and the median increases to 4 if we consider people younger than 30 years old. These results confirm once again the ease with which our algorithm can  find beautiful alternatives.

\subsubsection{Qualitative Results}
\begin{figure}[!t]
\center
 \includegraphics[width=.99\columnwidth]{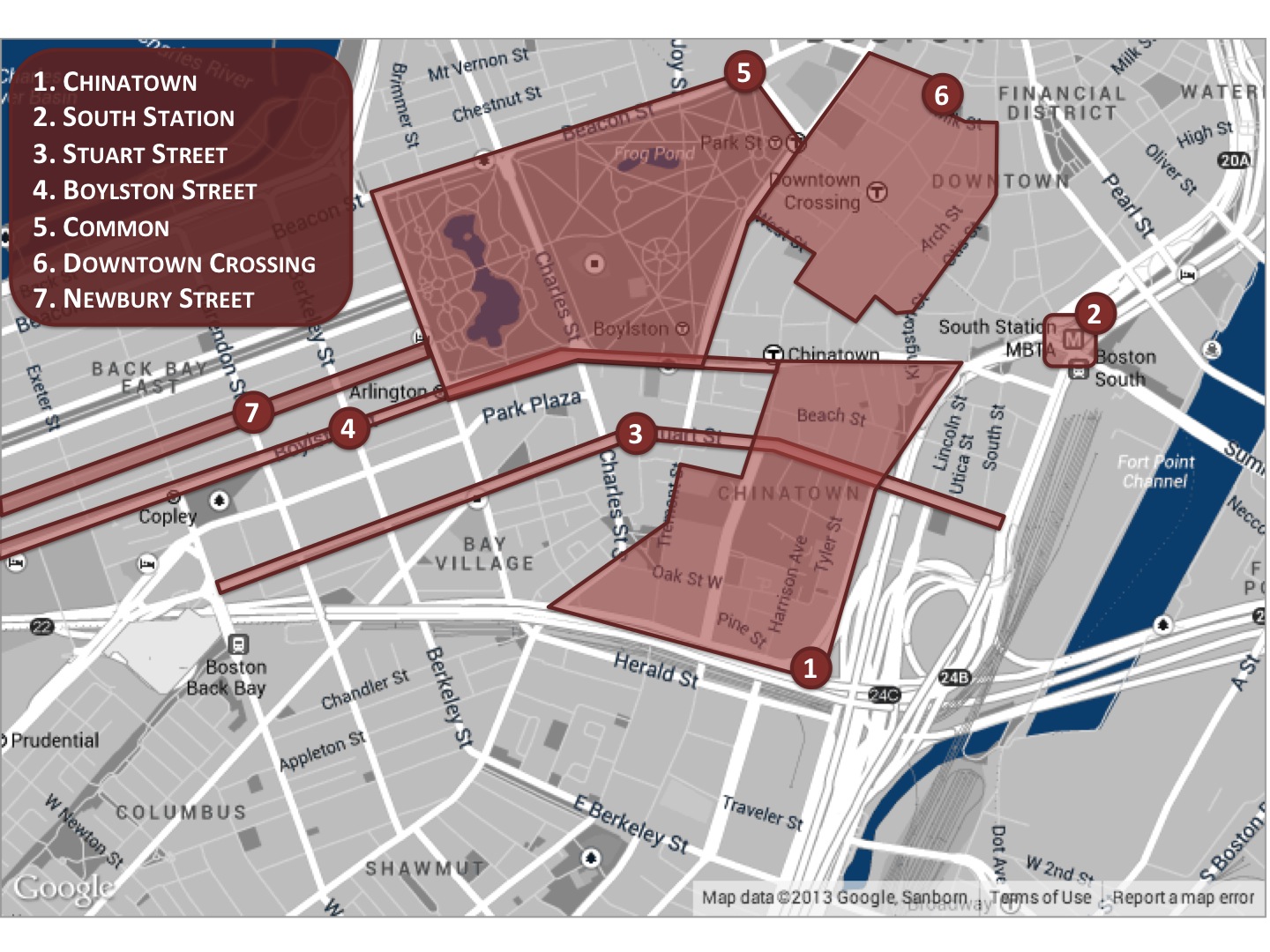}
 \footnotesize 
 \caption{Map of Boston with seven frequently mentioned places.}
 \label{fig:boston_map_cut}
\end{figure}

Our respondents see the Flickr-derived beautiful path positively. To ease illustration, the map in Figure~\ref{fig:boston_map_cut} shows the places our participants  frequently mentioned. The most positive feedback for the beautiful path is: ``You will see almost the entire city of Boston in shorter period of time''. By contrast,  the shortest path receives negative and neutral feedbacks. The most negative one mentions that a specific segment  ``is gross and no one needs to see it'', while the most positive feedback is neutral at best: ``It's an interesting and attractive walk \ldots but I wouldn't use the word  beautiful''. What is considered positive/negative in Boston is similar to what we found in London. On the positive side, respondents repeatedly mention green spaces: ``Boylston is really nice and they have a lot of shopping. The park is also gorgeous'', and another respondent adds ``green space,  manicure landscaping,  public activities  obviously all carefully designed to appeal to visitors as well as residents''. As for negative qualities, respondents cite the presence of cars (``It is heavily trafficked by cars in some areas'') and that of dirt (``Some places between South Station and Stuart are also rather dirty''). In a way similar to London, we also have mixed results for few locations: those are the ones that tend to be  used quite differently from day to night.
%. The most common reason was associated with how uses of those locations would change from day to night: ``It's nice to walk through Downtown Crossing because of the theaters but sometimes it can be dangerous at night''.

%There is limited scope of vision because the streets are relatively closed in and narrow here  

%
%beauty = 0.37 + 0.03 log(density) + 0.20 log(posemo)/c - 0.21 log.negative/c
%
%where
%* density is the column density of flickr pictures in the file
%* posemo is the column posemo in the file (i think it's the 70th column in your file, but you need to double check - i attached an xls file that might help)
%* log.negative = log(swear) + log(negemo) + log(anx) + log(sad) + log(anger)
%* c=#classified words in LIWC (this is the sum of all words)

\section{Discussion} 
\label{sec:06_discussion}

We are now ready to wrap up this work by  discussing some open questions.

\mbox{ } \\
\textbf{Scalability.} The two main steps of our approach are: 1) building the location graph with nodes and corresponding scores; and 2) finding the best path between source to destination in the graph. The complexity for the creation of the graph is negligible: it is linear with the number of nodes and has to be computed  just once for every city (even offline). The second step mainly consists of the Eppstein's algorithm, which is efficient as its complexity for computing a single path from source to destination is $O(e+n \cdot log(n)+m \cdot  log(m))$, where $n$ is the number of nodes, $e$ is the number of edges, and $m$ is the number of paths that are explored (e.g., in Boston our $m$ was just 700). More practically, a 24-core Intel server running Red Hat takes 51 millisecond  to compute the best path. 

\mbox{ } \\
\textbf{Personalization.} Perceptions of urban qualities such as happiness might well differ from one individual to another. In our user study, some respondents were happy to visit streets full of shops, while others cringed at the idea of facing a human mob of shoppers.
%The issue of individual preferences is best summarized by de Carteau in the chapter ``Walking in the City'' of his 1984 book ``The Practice of Everyday Life''~\cite{decerteau}. In it, he exposes the contradiction  between the objective methods of recording journeys   across the city by psychogeographers and the subjective natures of the individual histories they reflect.
Walking can be monitored and mapped, but as psycho-geographers observed, ``[walks on a map] can, in itself, never capture the personal histories that underlie them\ldots  \mbox{ } A map can never accurately capture the lives of those individuals whose journeys it sets out to trace, for in the process individuality is inevitably flattened out and reduced to points on a chart''~\cite{coverley2006psychogeography}. Personalization approaches might partly account for the subjectivity of  urban experiences  by, for example, tailoring recommended paths to a user's past visits~\cite{noulas:2012:RWA,Saez12}.

\mbox{ } \\
\textbf{Limited Spatial Representation.} To make our problem  computationally tractable, we had to model the city's spatial configuration in simple ways. We opted for dividing the city grid  into several cells, and those cells then represented nodes of our location graph. The finest unit of analysis consequently became the cell or, more specifically, its centroid. This means that we can capture neither whatever happens inside a cell (e.g., presence of a crime-infested council estate) nor the nature of the link that connects the centroids of two neighboring cells (e.g., crossing a large street full of cars).
%In this situation, we should pay attention to the potential consequences of the ecological fallacy, which refers to the case of considering each cell to be homogenous: everything inside it is considered to be the same.
More effective spatial representations are thus in order. One could, for example, resort to Space Syntax~\cite{social89}, a set of techniques for describing the spatial patterns produced by buildings and towns. These techniques would account for aspects we have so far left out from our analysis, including walkability, which is considered to be one of the most salient factors that make urban life thrive~\cite{speck2012walkable}.

\mbox{ } \\
\textbf{Salient City Pictures.} Our predictive models work on input of metadata associated with Flickr pictures. As we have not done any particular filtering, not all pictures are salient representations of the places in which they have been geographically tagged. Thus it might be useful to use existing techniques (e.g., computer vision algorithms) to determine the extent to which a picture represents its associated urban location.

\mbox{ } \\
\textbf{Limited Contextual Representation.}  Our study participants gave feedbacks that could be explored in future work. They, for example,  had fascinating insights related to weather conditions and  temporal aspects. One said:  ``The area around St Paul's can be very lovely. It depends rather on the time of day. At weekends when the City is quiet this will be nice. At busier times it could be manic''.  Future work might go into studying temporal dynamics at different levels:  time of the day (day \emph{vs.} night), day of the week (e.g., Saturday \emph{vs.} Monday), time of the year (e.g., different seasons). It would be also interesting to see how those dynamics change depending on, for example, weather conditions.

\mbox{ } \\
\textbf{Beyond route recommendations.} To offer an engaging user experience, designers have to build applications that go beyond just showing paths on a map. They could, for example, show the main points of interest along the path~\cite{DeChoudhury10automatic,ElAli13photographer}. Our survey respondents mentioned that historical elements contribute to a pleasant urban experience. A mobile app called ``NYPL Time Traveller'' has recently integrated the historical photographic collection of the New York Public Library with Foursquare, displaying historical NYC pictures related to the places in which they check in~\cite{foursquareblog}. Also, our respondents associated the concept of happiness not only with historical memories but also with their own personal stories. A respondent said: ``I would have a lot to say on the way about buildings, history, events and people I met along this path''. Therefore, one could also allow individuals to record their memories associated with a specific place and to show these memories back to them when physically revisiting that place, in a way similar to  StoryPlace.me or to the artistic project behind ``Map your Manhattan''~\cite{cooper2013mapping}.

\section{Conclusion}

Our goal has been to propose ways of recommending emotionally pleasing routes in the city. We have ascertained whether we met that goal in two steps: validation and evaluation. First, we have validated that our proposed route variations actually recommend what they are meant to (e.g., paths with highest perceived happiness), and they do so adding just a few extra walking minutes compared to the shortest routes. 

A user study involving 30 participants in London has shown that people perceive those recommendations the way we expect them to. Their quantitative assessments and qualitative insights all confirmed the results learned in the validation step.%, and that speaks to the external validity of our results. 

Finally, we have shown that we do not necessarily need to collect crowd-sourced ratings for every new city. Reasonable proxies can be computed from Flickr metadata. We did so for the city of Boston and ascertained its effectiveness with as many as 54 participants.

Viewing a path (no matter how familiar it is) does not capture the full affective experience of `being there'. In the future, we will build upon the analysis presented here  by designing a mobile application and testing it in the wild across different cities in Europe and USA.

\section{Acknowledgments}
We thank Adam Barwell for his important role in building the crowdsourcing site; Jon Crowcroft of Cambridge and the whole NetOS for continuous support; Abdo El Ali for his feedbacks on earlier versions of the user evaluation; friends at MIT, Northeastern, and Harvard  who helped with the survey; and Henriette Cramer, Giovanni Quattrone, and Olivier Van Laere  for their useful feedbacks. This work is supported by the SocialSensor FP7 project, partially funded by the EC under contract number 287975.

\balance

\small
\bibliographystyle{abbrv}
\bibliography{sigproc}

\balancecolumns

\end{document}